\documentclass{elsart} \usepackage{amssymb,epsfig}

\usepackage{graphics}
 
\def\be{\begin{equation}} \def\ee{\end{equation}}
\def\bq{\begin{eqnarray}} \def\eq{\end{eqnarray}}
\def\p{\bullet}

\begin{document}
\begin{frontmatter}

\title{ Color Superconductivity in Compact Stars and Gamma Ray Bursts}
\author{A. Drago$^a$, A. Lavagno$^b$, G. Pagliara$^a$}
\address{$^a$Dipartimento di Fisica, Universit{\`a} di Ferrara and
INFN, Sezione di Ferrara, 44100 Ferrara, Italy}
\address{$^b$Dipartimento di Fisica, Politecnico di Torino and INFN,
Sezione di Torino, 10129 Torino, Italy}

\date{\today } \maketitle

\begin{abstract} 
We study the effects of color superconductivity on the structure and
formation of compact stars. We show that it is possible to satisfy
most of recent observational boundaries on masses and radii if a
diquark condensate forms in a hybrid or a quark star.  Moreover, we
find that a huge amount of energy, of the order of $10^{53}$ erg, can
be released in the conversion from a (metastable) hadronic star into a
(stable) hybrid or quark star, if the presence of a color
superconducting phase is taken into account. Accordingly to the
scenario proposed in Astrophys.J.586(2003)1250, the
energy released in this conversion can power a Gamma Ray Burst. This
mechanism can explain the recent observations indicating a delay, of
the order of days or years, between a few Supernova explosions and the
subsequent Gamma Ray Burst.

\end{abstract}
\end{frontmatter}

\section{Introduction}

The new accumulating data from X-ray satellites provide important
information on the structure and formation of compact stellar
objects. Concerning the structure, the new data fix rather stringent
constraints on the mass and the radius of a compact star. These data
are at first sight difficult to interpret in a unique and
self-consistent theoretical scenario, since some of the observations
are indicating rather small radii and other observations are
indicating large values for the mass of the star.

Concerning the formation scenario, crucial information are provided by
the very recent observations of Gamma-Ray Bursts (GRB), indicating the
possibility that some of the GRBs are associated with a previous
Supernova explosion, with a delay between the first and the second
explosion of the order of days or years \cite{Amati00,Reeves02}.
These observations could be explained associating the second explosion
with the conversion of a (metastable) hadronic star (HS) into a more
stable stellar object made at least in part of deconfined quark matter
(QM). In the scenario proposed in Ref.\cite{noiapj} (see also references
therein), the HS can be metastable due to the presence of a
non-vanishing surface tension at the interface separating hadronic
matter (HM) from QM.  The nucleation time (i.e. the time
to form a critical-size drop of quark matter) can be extremely long if
the mass of the star is small.  Via mass accretion the nucleation time
can be dramatically reduced and the star is finally converted into the
stable configuration.

In recent years, many theoretical works have investigated the possible
formation of a diquark condensate in quark matter, at densities
reachable in the core of a compact star \cite{alf1,alf5,schafer}. The
formation of this condensate can deeply modify the structure of the
star \cite{alf4,baldo1,blas1}.

In this Letter we show that it is possible to satisfy the existing
boundaries on mass and radius of a compact stellar object if a diquark
condensate forms in a Hybrid Star (HyS) or a Quark Star (QS).
Moreover, the formation of diquark condensate can significantly
increase the energy released in the conversion from a purely HS into a
more stable star containing deconfined QM.

\section{Equation of state of beta-stable matter}

The EOS appropriate to the description of a compact star has to
satisfy beta-stability conditions. Moreover two charges are conserved,
the baryonic and the electric one.  These conditions need to be
satisfied in the hadronic, in the quark and in the mixed phase.
Although electric charge neutrality and beta-stability are easy to
impose for non-interacting quark matter, these conditions are highly
non-trivial when a diquark condensate can develop.  

Concerning the hadronic phase we use the relativistic non-linear
Glendenning-Moszkowski model (GM1-GM3) \cite{glen2}.  
At very low density we have used the Negele-Vautherin \cite{negele} and
the Baym-Pethick-Sutherland \cite{baym} EOS.
For the quark
matter phase we adopt a MIT-bag like model in which the formation of a
diquark condensate is taken into account in a simple and effective way
which will be described below.  To connect the two phases of our EOS,
we impose Gibbs equilibrium conditions.  When the Gibbs conditions are
applied in presence of more than one conserved charge, the technique
developed by Glendenning has to be adopted \cite{glen4,glen5} and the
pressure need not to be constant in the mixed phase.  Therefore a
finite volume of the star can be occupied by the mixed phase, what is
crucial for the stability of the star.

It is widely accepted that the Color-Flavor Locking phase (CFL) is the
real ground state of QCD at asymptoticly large densities but, at the
scales which are involved in a compact star, there is yet uncertainty
about the presence of this phase and in particular about the
transition from superconducting QM to HM. There are different possible
scenarios depending on the value of the strange quark mass. A direct
transition from CFL to HM is possible for a small value of $m_s$.
Alternatively, an intermediate window of Crystalline Color
Superconductivity phase can develop \cite{raj1,raj3} or a phase can be
formed in which the Cooper pairing involves only quarks of the same
flavor \cite{alf3}. A pure two flavor color superconductivity (2SC)
phase is ruled out due to the high free energy cost resulting from the
requirement of color and electric neutrality \cite{alf3,steiner1}.  A
mixed 2SC-CFL phase could exist but it seems to be unstable due to
Coulomb and surface effects \cite{buba1}.  In our model the CFL phase
is connected directly to HM through a first order phase transition,
what seems to be consistent with the use of a small value for $m_s$,
of the order of 150 MeV.

In this Letter we are interested in the bulk properties of a compact
star.  We use therefore a schematic model which takes into account in
a simple and effective way the main characteristics of the EOS of
quark matter in the presence of diquark condensation. The main aim of
our work is to discuss the dependence of the structure of the star on
the numerical value of three crucial parameters, namely the height and
position of the maximum of the diquark gap and the value of the
pressure of the vacuum $B$ of the MIT bag model.

We adopt the scheme proposed in Refs.\cite{alf4,lugo1} where the
thermodynamic potential is given by the sum of two contributions. The
first term corresponds to a ``fictional'' state of unpaired quark
matter in which all quarks have a common Fermi momentum chosen to
minimize the thermodynamic potential. The other term is the binding
energy of the diquark condensate expanded up to order
$(\Delta/\mu)^2$.  In Ref.\cite{alf4} the gap is assumed to be
constant, independent on the chemical potential. In the present
calculation we consider a $\mu$ dependent gap resulting from the
solution of the gap equation.  We describe the superconducting phase
using the first work of Alford, Rajagopal and Wilczek in which the
Color-Flavor Locking phase was introduced \cite{alf1} by considering
three massless flavors.  This approximation is a sensible one as long
as $m_s$ is small in comparison with the quark chemical potential
\cite{raj1}.  The quark-quark interaction is described by a NJL-like
Lagrangian with an effective coupling constant $K$ and a form factor
which mimics the asymptotic freedom of QCD. The form factor reads:
$ F(k)=\left( 1 +
\exp\Bigl[ {k-\Lambda \over w} \Bigr] \right)^{-1}\, , $ 
where $w$
and $\Lambda$ are free parameters of this model.

The CFL phase is characterized by the existence of two order
parameters $\Delta_s$ (singlet) and $\Delta_o$ (octet) which are the 
solutions of two
coupled gap equations.  The value of the coupling $K$ is fixed, as in
Ref.\cite{alf1}, by imposing that the NJL Lagrangian with the same $K$
gives, through chiral symmetry breaking at zero chemical potential, a
reasonable value for the chiral gap $\Delta_{\chi}$. This procedure in
turn fixes the maximum of the superconducting gap. In this work we use
values for $K$ giving a $\Delta_{\chi}$ ranging from 300 to 600 MeV
and a corresponding maximum of the ``effective gap'' 
$\Delta(\mu)=F^2(\mu)\sqrt{[8(\Delta_o(\mu))^2+(\Delta_s(\mu))^2]/12} $
varying from 70 to 150 MeV.  This definition of $\Delta$ corresponds
to the gap used in Ref.\cite{alf4} if $\Delta_s=2\vert\Delta_o\vert$,
as assumed in that work.  In Fig.1 we display the ``effective gap'' as
a function of the chemical potential.  In our calculation we have not
really solved self-consistently the coupled equations for the chiral
and the superconducting gap. The result of microscopic calculations,
like the ones of Refs.\cite{steiner1,buba1} indicates that the two
gaps are mutually exclusive. In particular, the superconducting gap is
suppressed at low $\mu$.  Therefore we do not consider realistic the
parameters corresponding to gaps $\Delta_1$ and $\Delta_2$, and they
are mainly introduced to illustrate the effects connected with the
position $\mu_{max}$ of the maximum of the gap.

In our model, confinement is schematically described by introducing
the MIT bag constant $B$.  Moreover, the pressure and the energy density  
are modified by the contributions of the electrons, which are
necessary in the mixed phase\footnote{In the pure CFL phase the
contribution of electrons vanishes due to the electrical neutrality
enforced by the existence of the gap \cite{raj2}.}:
\bq
P&=&-\Omega_{CFL}(\mu)-B-\Omega^{electrons}(\mu_e)\\
E/V&=&\Omega_{CFL}(\mu)+\mu\rho+B+\Omega^{electrons}(\mu_e)+\mu_e\rho_e
\, , 
\eq 
where
\be
\Omega_{CFL}(\mu)={6\over \pi^2}\int_0^\nu k^2(k-\mu)\,{\mathrm d}k+
{3\over \pi^2}\int_0^\nu k^2(\sqrt{k^2+m_s^2}-\mu)\,{\mathrm d}k-
{3 \Delta^2 \mu^2\over \pi^2}
\ee
with
\be
\nu=2\mu-\sqrt{\mu^2+{m_s^2\over 3}}\, ,
\ee
and the quark density $\rho$ is calculated numerically
by deriving the thermodynamic potential respect to $\mu$.

In Fig.2 we show the EOS with and without color superconductivity.
The effect of the gap is to increase the pressure of paired QM respect
to the unpaired QM at a fixed chemical potential.  This extra pressure
reduces the values of the critical densities.  Comparing the curves
for $\Delta=0$ and $\Delta\neq0$ we can see that the CFL EOS is softer
than the unpaired QM EOS at low density and stiffer at high
density. This will have important consequences in the M-R curves.

\begin{figure}
\label{gap}
\parbox{6cm}{ \scalebox{0.6}{
\includegraphics*[-40,430][560,730]{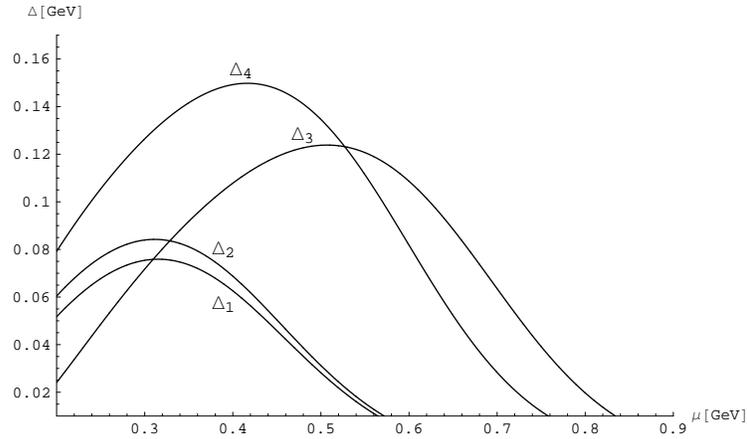}} }

\parbox{14cm}{
\caption{Gap as function of the chemical potential, for four different
parameter sets.}}
\vspace{1cm}
\end{figure}

\begin{figure}
\label{eos}
\parbox{6cm}{ \scalebox{0.6}{
\includegraphics*[-40,430][550,730]{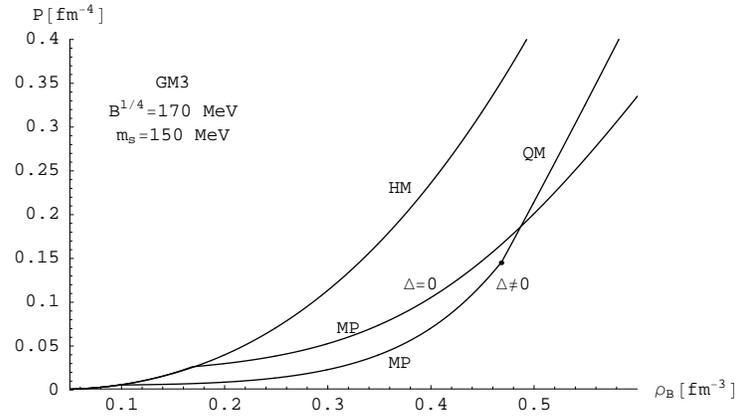}} }

\parbox{14cm}{
\caption{Pressure versus baryonic density. HM indicates a purely
hadronic EOS, MP a mixed-phase of hadrons and quarks and QM pure quark
matter. The effect of a non-vanishing superconducting gap is
displayed.  }}
\vspace{1cm}
\end{figure}

\begin{figure}
\label{masseraggi}
\parbox{6cm}{ \scalebox{0.7}{
\includegraphics*[25,410][550,800]{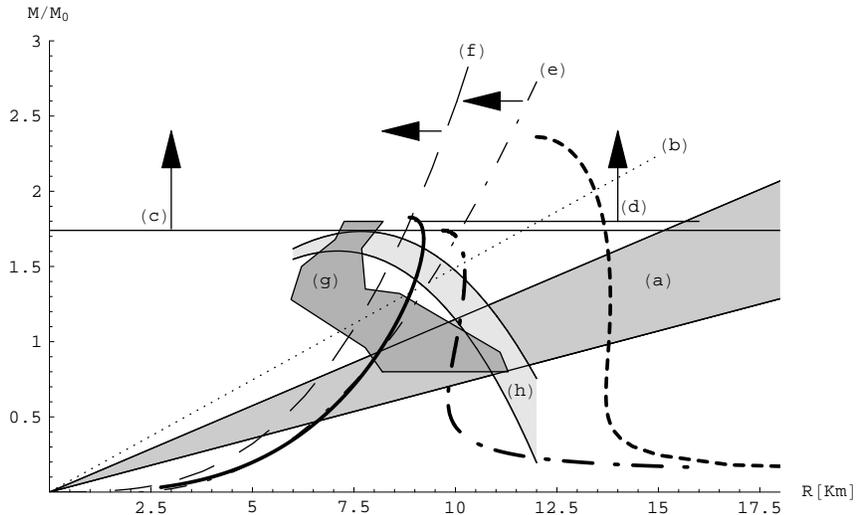}} }

\parbox{14cm}{
\caption{Mass-radius plane with observational limits and a few representative
theoretical curves: thick solid line indicates CFL quark stars, 
thick dot-dashed line
CFL hybrid stars, thick-dashed line hadronic stars (see text). The
observational limits come from: (a)\,Sanwal et al. 2002 \cite{sanwal},
(b)\, Cottam et al. 2002 \cite{cottam}, (c)\, Quaintrell et al. 2003
\cite{quaint}, (d)\, Heinke et al. 2003 \cite{heinke}, (e),(g)\, Dey
et al. 1998 \cite{dey}, (f)\, Li et al. 1999 \cite{li}, (h)\, Burwitz
et al. 2002 \cite{burwitz}.  }}
\vspace{1cm}
\end{figure}

\section{Masses and radii of compact stellar objects}

In Fig.3 we have collected most of the analysis of data from X-ray
satellites, concerning masses and radii of compact stellar objects
\cite{sanwal,cottam,quaint,heinke,dey,li,burwitz}.  Most of the data
have been obtained very recently due to the technological progresses
in the field of X-ray detectors. Although some (or all) of the data
analysis are controversial, since they depend on specific assumptions
on the structure of the X-ray source, we do think that these
observational results deserve to be carefully discussed.

Observing Fig.3, we notice that the constraints coming from a few data
sets (labeled ``e'', ``f''\footnote{A very recent reanalysis of the
data of the pulsar SAX J1808.4-3658, discussed in Ref. \cite{li}, seems
to indicate slightly larger radii, of the order of 9-10 km for a star
having a mass of 1.4-1.5 $M_\odot$ \cite{Poutanen}.}  ``g'' and maybe
also constraint ``h''\footnote{The data at the origin of constraint
``h'' have been discussed in many recent papers. In Ref.\cite{zane} an
indication for an even more compact stellar object can be found.
Anyway, the so-called thermal radius obtained in these analysis could
be significantly smaller than the total radius of the star.})
indicate rather unambiguously the existence of very compact stellar
objects, having a radius smaller than $\sim 10$ km.  At the contrary,
at least in one case (``a'' in the figure), the analysis of the data
suggests the existence of stellar objects having radii of the order of
12 km or larger, if their mass is of the order of 1.4 $M_\odot$.  In
this analysis one has also to take into account that it is difficult
from an astrophysical viewpoint to generate compact stellar objects
having a mass of the order of one solar mass or smaller.  Therefore
the most likely interpretation of constraint ``a'' is that the
corresponding stellar object does not belong to the same class of
objects which have a radius smaller than $\sim 10$ km. Concerning
constraint ``b'', its interpretation is less clear, since it can be
satisfied both with a very compact star or with a star having a larger
radius.  The apparent contradiction between the constraints ``e'',
``f'', ``g'' and the constraint ``a'' can be easily accommodated in our
scheme, since it can be the signal of the existence of metastable
purely hadronic stars which can collapse into a stable configuration
when deconfined quark matter forms inside the star. In the next
Section we will discuss the possible relation between this transition
and at least some GRBs.

Finally, constraints (``c'' and ``d'') do not provide stringent limits
on the radius of the star, but they put strong constraints on the
lower value of its mass.  Constraints ``c'' and ``d'' are very
important, since it is in general not easy to obtain solutions of the
Tolman-Oppenheimer-Volkoff equation having both large masses and very
small radii. As we will see, the existence of an energy gap associated
with the diquark condensate helps in circumventing this difficulty,
since the effect of the gap is to increase the maximum mass of QSs or
of HySs having a huge content of pure quark matter, as shown in Fig.4.

In the upper panel of Fig.4 we show that a CFL HyS has a smaller
radius respect to an unpaired QM HyS, if the mass of the star is
smaller than $\sim$ 1.35 $M_\odot$.  This can be explained observing
the EOSs shown in Fig.2.  For a low mass star, the central density
lies in the region of the $P-\rho$ plane where the CFL EOS is softer
than the unpaired QM EOS and therefore the radius of the star is
smaller. For a large mass star, on the contrary, the central density
is in the region where the CFL EOS is stiffer than unpaired QM EOS and
therefore the radius of the star is larger. It is also worth remarking
that if the value of the gap is increased the amount of QM in the star
also increases.  Therefore, for large values of the gap, heavy HySs
have a shape more and more similar to the shape of pure QSs, which are
finally obtained by a further increase of the value of the
gap\footnote{Notice that the existence of a large CFL gap is not
constrained by the traditional argument concerning the stability of Fe
against decay into two-flavor QM.}  (see also the lower panel of
Fig.4).  All these results are totally consistent with the ones
obtained in Ref.\cite{alf4}.

Concerning the value of the chemical potential $\mu_{max}$, which
corresponds to the maximum of the gap, we get relevant modifications
to the mass-radius relation if $0.3$ GeV $\lesssim\mu_{max}\lesssim
0.6$ GeV. For larger values of $\mu_{max}$ the effect of the gap is
negligible. Low values of $\mu_{max}$, of the order of 0.3 GeV,
correspond to interesting stellar configurations, but are difficult to
justify at the light of results like the ones presented in
Ref.\cite{buba1} (see discussion at the end of Sec.2). In the upper
panel of Fig.4 we display results for three values of $\mu_{max}$,
whose corresponding gaps are shown in Fig.1. We also
show that, using $B^{1/4}$ = 170 MeV, only a big value of the gap,
located at a not too large density ($\Delta_4$) allows the formation
of a QS, while for the other gaps HySs are obtained.

In Fig.3 we show a few theoretical M-R relations which correspond to
the scenario we are proposing. More precisely, we show a 
thick-dashed line corresponding to HSs (GM1), a thick dot-dashed line 
corresponding to HySs
(GM1, $B^{1/4}=170$ MeV, $\Delta_2$) and a thick solid line
corresponding to QSs (GM1, $B^{1/4}=170$ MeV, $\Delta_4$).  
Both the HyS and the QS lines can
satisfy essentially all the constrains derived from observations
(concerning the constraint ``f'' see footnote (2)).  Concerning the
constraint ``a'', it is probably better satisfied by the HS line than
by the HyS or QS lines, which would give stars having a mass smaller
than $\sim 1.2 M_\odot$.  In conclusion, in our scheme most of the
compact stars are either HySs or QSs having a mass in the range
$1.2-1.8 M_\odot$ and a radius $\sim 8.5-10$ km.  Metastable HS can
exist. As we will see in the next section their mass is probably
smaller than $\sim 1.3 M_\odot$.

\begin{figure}
\label{mr}
\parbox{6cm}{ \scalebox{0.6}{
\includegraphics*[-50,140][590,900]{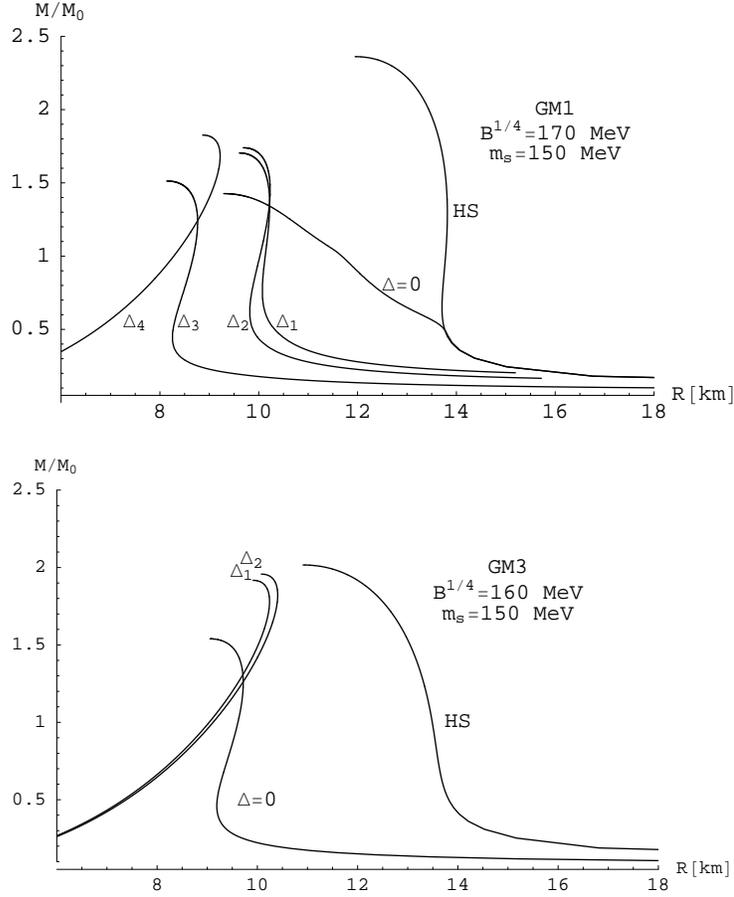}} }

\parbox{14cm}{
\caption{A few theoretical mass-radius relations are shown. HS indicates purely
hadronic stars. The other lines correspond either to hybrid
or quark stars, depending on the value of the gap for a given value of $B$.}}
\vspace{1cm}
\end{figure}

\section{Nucleation time and energy released}

In the model we are discussing the formation of QSs or HySs is due to
the conversion of a purely HS into a more compact star in which
deconfined QM is present. An HS can be metastable if a non-vanishing
surface tension is present at the interface between HM and QM.  The
process of quark deconfinement can be a powerful source for GRBs and
it can also explain the delay between a supernova explosion and the
subsequent GRB observed in a few cases
\cite{Amati00,Reeves02}\footnote{A possible mechanism explaining these
GRBs is the supranova model \cite{vietri}. In this model, the GRB is
the result of the collapse to a black hole of a supramassive fast
rotating NS, as it loses angular momentum.  According to this model NS
is produced in the Supernova explosion preceding the GRB event.  The
initial baryonic mass of the NS is assumed to be above the maximum
baryonic mass for non-rotating configurations. As noticed in
Ref. \cite{bottcher} in these collapse too much baryonic material is
ejected and thus the energy output is expected to be too small to
produce GRBs.  Moreover, the supranova model seems to produce GRBs
which are too short compared with the observed durations.}.  In the
scenario proposed in Ref.\cite{noiapj}, the central density of a pure
HS increases, due to spin down or mass accretion, until its value
approaches the deconfinement critical density. At this point a
spherical virtual drop of QM can form. The potential energy for
fluctuations of the drop radius $R$ has the form \cite{lif}:
\be
U(R)={4 \over 3} \pi R^3 n_q (\mu_q-\mu_h)+4 \pi\sigma R^2 + 8
\pi\gamma R 
\ee 
where $n_q$ is the quark baryon density, $\mu_h$ and $\mu_q$ are the
hadronic and quark chemical potentials, all computed at a fixed
pressure $P$, and $\sigma$ is the surface tension for the interface
separating quarks from hadrons. Finally, the term containing $\gamma$
is the so called curvature energy.  For $\sigma$ we use standard
values from 10 to 40 MeV/fm$^2$ and we assume that it takes into
account, in a effective way, also the curvature energy. The value of
$\sigma$ was estimated in Ref.\cite{jaffe} to be $\sim 10$
MeV/fm$^2$. Values for $\sigma$ larger than $\sim 30$ MeV/fm$^2$ are
probably not useful at the light of the result of
Refs.\cite{voskresensky,alf2}.

To compute the time needed to form a bubble of quarks having a radius
larger than the critical one, we use the technique of quantum tunneling
nucleation.  We can assume that the temperature has no effect in our
scheme: for values of $B^{1/4}\sim160-180$ MeV, which we use in this
Letter, the critical density $\rho_1$ separating pure HM from mixed
phase is larger than $4\rho_0$ for $Z/A\sim0.3$, i.e. for an isospin
fraction typical of a newly formed and hot proto-neutron star
\cite{ditoro}. This critical density typically exceeds the central
density of hot and not too massive stars.  Therefore the mixed phase
can form only when the star has deleptonized and its temperature has
dropped down to a few MeV \cite{Pons}.  When the temperature is so
low, only quantum tunneling is a practicable mechanism.  The calculation
proceed in the usual way: after the computation (in WKB approximation)
of the ground state energy $E_0$ and of the oscillation frequency
$\nu_0$ of the virtual QM drop in the potential well $U(R)$, it is
possible to calculate in a relativistic frame the probability of
tunneling as \cite{iida}:
\be p_0=\exp [-{A(E_0)\over \hbar}] 
\ee
where 
\be 
A(E)={2}\int_{R_-}^{R_+} dR \sqrt{[2 M(R)+E-U(R)][U(R)-E]}\,.  
\ee 
Here $R_\pm$ are the classical turning points and 
\be
M(R)=4\pi\rho_h\left(1-{n_q\over n_h}\right)^2 R^3 \,\,\,\,\, , 
\ee
$\rho_h$ being the hadronic energy density and $n_h$, $n_q$ are the
baryonic densities at a same and given pressure in the hadronic and
quark phase, respectively.  The nucleation time is then equal to \be
\tau = (\nu_0 p_0 N_c)^{-1}\, ,
\label{delay}
\ee where $N_c$ is the number of centers of droplet formation in the
star, and it is of the order of $10^{48}$ \cite{iida}.  In the
calculation of nucleation times we neglect the effects of color
superconductivity.  We assume that the CFL gap cannot form until the
radius of the quark drop has increased enough and therefore the energy
released in the pairing process has not to be taken into account when
computing the nucleation time.  A support to our assumption can be
found in Ref.\cite{amor1}, where the authors have investigated finite
size effects on the formation of a 2SC gap, projecting onto states of
defined baryon number and onto color singlets.  If the radius of the
quark nugget is smaller than a critical length (of the order of 1.5--2
fm in their case) the magnitude of gap is drastically reduced.  A
similar calculation for the CFL phase has not yet been done, but it is
reasonable to assume that due to the ``locking'' between color and
flavor in the CFL phase the color projection will yield even larger
effects.

Let us recall once again the astrophysical scenario we have in
mind. In a few cases a delay of the order of days or years between the
Supernova explosion and the subsequent GRB have been postulated to
explain the astrophysical data on the GRBs.  In the scheme we are
discussing, this delay is due to the formation of a metastable HS
having a relatively small mass. The nucleation time, computed using
Eq. (\ref{delay}), can be extremely long if the mass of the metastable
star is small enough. Via mass accretion the nucleation time can be
reduced from values of the order of the age of the universe down to a
value of the order of days or years. We can therefore determine the
critical mass $M_{cr}$ of the metastable HS for which the nucleation
time corresponds to a fixed small value (1 year in Table 1).

\begin{table}[ht]
\begin{center}
\tabcolsep=0.4\tabcolsep 
\begin{tabular}{ccccccccc}
\hline
\hline
Hadronic         &
$B^{1/4}$        & 
$\sigma$         & 
$M_{cr}/M_\odot$ & 
$\Delta E$  & 
$\Delta E$  &  
$\Delta E$  & 
$\Delta E$  &
$\Delta E$  \\
Model        &
[MeV]        &
[MeV/fm$^2$] & 
             & 
$\Delta=0$   &  
$\Delta_1$   &   
$\Delta_2$   &  
$\Delta_3$   &
$\Delta_4$   \\        

\hline
GM3 & $160$  & $20$ & $0.69$  & $20$ &  $65^\p$    & $69^\p$  & $76^\p$  & $148^\p$ \\
GM3 & $160$  & $30$ & $0.91$  & $32$ &  $90^\p$&   $95^\p$     &  $106^\p$ & $196^\p$  \\
GM3 & $160$  & $40$ & $1.00$  & $38$ &  $100^\p$&  $105^\p$ &  $119^\p$ & $216^\p$  \\
\hline
GM3 & $170$  & $10$ & $1.12$  & $0$  &  $34$   &  $40$   & $68$ & $162^\p$  \\
GM3 & $170$  & $20$ & $1.26$  & $4$  &  $44$   &  $50$   & $86$ & $185^\p$   \\
GM3 & $170$  & $30$ & $1.39$  & $11$ &  $53$   &  $60$   & $104$ & $207^\p$   \\
GM3 & $170$  & $40$ & $1.49$  & BH   &  $62$   &  $68$   & $120$ & $224^\p$   \\
\hline
GM3 & $180$  & $10$ & $1.55$  & BH  &  11   &  $13$   & BH & --  \\
GM3 & $180$  & $20$ & $1.61$  & BH  &  BH   &  22   &   BH   & --  \\
GM3 & $180$  & $30$ & $1.67$  & BH  &  BH   &  BH   & BH   & --   \\
\hline
GM1 & $160$  & $10$ & $0.45$  & $11$ &  $41^\p$ &   $44^\p$ &  $47^\p$ & $96^\p$   \\
GM1 & $160$  & $20$ & $0.72$  & $28$ &  $75^\p$ &   $79^\p$ &  $86^\p$ & $160^\p$  \\
GM1 & $160$  & $30$ & $0.96$  & $48$ &  $108^\p$ &   $114^\p$ &  $127^\p$ & $220^\p$   \\
GM1 & $160$  & $40$ & $1.18$  & $72$ &  $142^\p$ &  $148^\p$ &  $166^\p$ & $276^\p$    \\
\hline
GM1 & $170$  & $10$ & $1.17$  & $18$ &  $59$ &  $65$  &  $96$ & $191^\p$  \\
GM1 & $170$  & $20$ & $1.33$  & $33$ &  $79$ &  $85$  &  $124$ & $226^\p$ \\
GM1 & $170$  & $30$ & $1.45$  & $50$ &  $96$ &  $103$  &  $150$ & $254^\p$  \\
GM1 & $170$  & $40$ & $1.60$  & BH  & $122$ &  $128$  & BH & $290^\p$   \\
\hline
GM1 & $180$  & $10$ & $1.63$  & BH  &  BH     &  72   & BH & --    \\
GM1 & $180$  & $20$ & $1.72$  & BH  &  BH     &  BH   & BH & --    \\
GM1 & $180$  & $30$ & $1.79$  & BH  &  BH     &  BH   & BH & --     \\

\hline \hline
\end{tabular}
\vspace{1cm}
\caption{Energy released $\Delta E$ (measured in foe=10$^{51}$ erg) in
the conversion to hybrid or quark star (labeled with a $\p$), for
various sets of model parameters, assuming the hadronic star mean
life-time $\tau=1$ yr (see text). $M_{cr}$ is the gravitational mass
of the hadronic star at which the transition takes place, for fixed
values of the surface tension $\sigma$ and of the mean life-time
$\tau$. BH indicates that the hadronic star collapses to a Black Hole.
We indicate with a dash (--) situations in which the Gibbs
construction does not provide a mechanically stable EOS.}
\end{center}
\label{}
\end{table}

In Table 1 we show the value of $M_{cr}$ for various sets of model
parameters.  In the conversion process from a metastable HS into an
HyS or a QS a huge amount of energy $\Delta E$ is released.  $\Delta
E$ is the difference between the gravitational mass of the metastable
HS and that of the final HyS or QS having the same baryonic mass.  We
see in the Table that the formation of a CFL phase allows to obtain
values for $\Delta E$ which are one order of magnitude larger
than the corresponding $\Delta E$ of the unpaired QM case
($\Delta=0$).  Moreover, we can observe that $\Delta E$ depends both on
magnitude and position of the gap.

Finally let us comment on the dependence of the results of Table 1 on
the value of $B$. As we can see, if $B^{1/4}<160$ MeV the value of
$M_{cr}$ is very small and it is unlikely that a metastable HS having
a mass $M \le M_{cr}$ can be obtained from a Supernova explosion. In
that case all compact stars would be either HySs or QSs. If, at the
contrary, $B^{1/4}>170$ MeV, the value of $M_{cr}$ is so large that a
compact star can be only a (metastable) HS since, after the conversion
from HM into QM, the HS collapses into a Black Hole (BH). Therefore,
only if 160 MeV $\lesssim B^{1/4}\lesssim$ 170 MeV a GRB can be
generated within the mechanism we are proposing.

\section{Conclusions}
We have studied the effect of color superconductivity on the EOS of quark
matter and on the mass-radius relation for hybrid and quark
stars. Comparing the theoretical curves with recent analysis of
observational data, we find that color superconductivity is a crucial
ingredient in order to satisfy all the constraints coming from
observations. The most difficult problem posed by the astrophysical
data is the indication of the existence of stars which are both very
compact (R $\lesssim$ 9--10 km) and rather massive ($M\gtrsim 1.7
M_\odot$).  We can satisfy these constraints either with hybrid or
quark stars.  In particular, concerning hybrid stars, the gap
increases significantly the maximum mass of the stable configuration,
while keeping the corresponding radius $\lesssim$ 10 km. These
findings are in agreement with the ones of Ref.\cite{alf4}, where a
chemical potential independent gap was used. Anyway, we have shown in
this Letter that, to obtain very large masses for hybrid stars, the
maximum of the superconducting gap need to be located at a value of
the chemical potential which is probably too small.

The superconducting gap affects also deeply the energy released in the
conversion from hadronic star into hybrid or quark star.  To explain
recent observations indicating a delay between a Supernova explosion
and the subsequent Gamma Ray Burst \cite{Amati00,Reeves02}, in 
Ref.\cite{noiapj} it has been
proposed to associate the second explosion with the transition from a
metastable hadronic star to a stable star containing deconfined quark
matter. In this Letter we have shown that the energy released, 
which will power the Gamma Ray Burst, is significantly increased by the 
effect of the superconducting gap and it can reach a value of the order 
of $10^{53}$ erg.

To satisfy the constraints on masses and radii of the compact stellar
objects and to obtain a huge energy from the conversion of the
metastable hadronic star into a quark or hybrid star, rather stringent
limits on the parameter values have to be imposed.  More explicitly,
the pressure of the vacuum has to be in the range 160 MeV $\lesssim
B^{1/4}\lesssim$ 170 MeV and the value of the gap has to have a
maximum $\Delta\sim 0.15$ GeV at a chemical potential 0.4 GeV
$\lesssim\mu_{max}\lesssim$ 0.6 GeV. All these parameters are
compatible with the results both of hadronic physics calculations
and of microscopic studies of superconducting quark matter.
Finally, let us remark that for values of $B$ in the indicated range, 
hybrid stars are obtained for unpaired quark matter, while, in most cases, the
formation of the superconducting gap yields quark star configurations.

Very recently, a new candidate for a delayed Supernova -- Gamma Ray
Burst association, with an estimated delay of the order of months, has
been proposed in Ref.\cite{butler}. If confirmed, this observation
would constitute an important support to models like the one we have
suggested.

\bigskip

It is a pleasure to thank I.~Bombaci, G.~Fiorentini, F.~Frontera,
D.~Guetta and S.~Zane for many useful discussions.

\end{document}